\documentclass[conference,9pt]{IEEEtran}
\IEEEoverridecommandlockouts
\usepackage{algpseudocode}
\usepackage{amsmath}
\usepackage{booktabs}
\usepackage{stfloats}
\usepackage[ruled, vlined]{algorithm2e}
\usepackage{tikz}
\usetikzlibrary{decorations.pathreplacing}
\usepackage{arydshln}
\usepackage{enumitem}
\usepackage{algpseudocode}
\usepackage{multirow}
\usepackage{epsfig}
\usepackage{multicol,lipsum,xparse}
\usepackage{caption}
\usepackage{subcaption}
\usepackage{stmaryrd}
\usepackage{hyperref}
\usepackage{tikz}
\usepackage{subcaption}

\usepackage{cleveref}

\usepackage{todonotes}
\usepackage{cite}
\usepackage{amsmath,amssymb,amsfonts}
\usepackage{graphicx}
\usepackage{textcomp}
\usepackage{xcolor}
\usepackage[a4paper, total={184mm,239mm}]{geometry}
\def\BibTeX{{\rm B\kern-.05em{\sc i\kern-.025em b}\kern-.08em
    T\kern-.1667em\lower.7ex\hbox{E}\kern-.125emX}}

\begin{document}
\pagenumbering{roman}

\title{RESPECT: \underline{R}einforcement Learning based \underline{E}dge \underline{S}cheduling on \underline{P}ipelined \underline{C}oral \underline{E}dge \underline{T}PUs}

\newcommand{\fixme}[1]{\textcolor{red}{\small [~#1~]}}
\author{\IEEEauthorblockN{Jiaqi Yin}
\IEEEauthorblockA{
\textit{University of Utah}\\
Salt Lake City, US \\
jiaqi.yin@utah.edu}
\and
\IEEEauthorblockN{Yingjie Li}
\IEEEauthorblockA{
\textit{University of Utah}\\
Salt Lake City, US \\
yingjie.li@utah.edu}
\and
\IEEEauthorblockN{Daniel Robinson}
\IEEEauthorblockA{
\textit{University of Utah}\\
Salt Lake City, US \\
u0714849@umail.utah.edu}
\and
\IEEEauthorblockN{Cunxi Yu}
\IEEEauthorblockA{
\textit{University of Utah}\\
Salt Lake City, US \\
cunxi.yu@utah.edu}
}



\newcommand*\circled[1]{\raisebox{.4pt}
                    {\tikz[baseline=(char.base)]{
            \node[shape=circle,draw,inner sep=1pt, style={fill=black, text=white}, scale=0.75] (char) {\textbf{#1}};}}}
\maketitle

\begin{abstract}
Deep neural networks (DNNs) have substantial computational and memory requirements, and the compilation of its computational graphs has a great impact on the performance of resource-constrained (e.g., computation, I/O, and memory-bound) edge computing systems. While efficient execution of their computational graph requires an effective scheduling algorithm, generating the optimal scheduling solution is a challenging \textit{NP-hard} problem. Furthermore, the complexity of scheduling DNN computational graphs will further increase on pipelined multi-core systems considering memory communication cost, as well as the increasing size of DNNs. Using the synthetic graph for the training dataset, this work presents a reinforcement learning (RL) based scheduling framework \textbf{RESPECT}, which learns the behaviors of optimal optimization algorithms and generates near-optimal scheduling results with short solving runtime overhead. Our framework has demonstrated up to $\sim2.5\times$ real-world on-chip inference runtime speedups over the commercial compiler with ten popular ImageNet models deployed on the physical Coral Edge TPUs system. Moreover, compared to the exact optimization methods, the proposed RL scheduling improves the scheduling optimization runtime by up to 683$\times$ speedups compared to the commercial compiler and matches the exact optimal solutions with up to 930$\times$ speedups. Finally, we perform a comprehensive generalizability test, which demonstrates RESPECT successfully imitates optimal solving behaviors from small synthetic graphs to large real-world DNNs computational graphs.

\end{abstract}

\section{Introduction}\label{sec:intro}



To efficiently utilize hardware platforms, scheduling algorithms implemented in deep learning compilers are critical in deploying these hyper-dimensional computationally-intensive workloads, and this is a classical \textit{NP-hard} combinatorial optimization (CO) problem. Mostly, vendor-specific libraries rely on hand-crafted domain-specific heuristics to optimize the executions, which trades the execution performance for scheduling solving time. Specifically, the limitations of the existing scheduling approach can be summarized as follows: \textbf{(1)} existing algorithms are either heuristics that lack in quality of optimization or exact/brute-force algorithms lacking in scalability \cite{zhang2013sdc,CoZh06:sdc,paulin1989force,yin2022exact}. The challenges for large-scale deep learning executions are still rising, particularly for edge devices; \textbf{(2)} hand-crafted heuristics can be efficient but the development process requires high engineering efforts and domain knowledge for compilation and hardware systems; \textbf{(3)} there have recently seen machine learning (ML) based frameworks for design automation \cite{yu2018developing,yu2019painting,ren2023machine} and compilation \cite{mao2019learning,chen2019deep,sheng2021deep}. However, they are either limited to certain graph structures/sizes or require online training or retraining \cite{mao2019learning,chen2019deep,sheng2021deep}. More importantly, existing ML-based approaches all rely on real-world data collection, which is time-consuming and limited in generalizability \cite{ustun2019lamda,ren2023machine,pal2022machine}.

This work presents RESPECT, a novel combinatorial reinforcement learning framework that imitates the behaviors of graph combinatorial algorithms for scheduling DNN computational graphs on edge devices. RESPECT aims to perform near-optimum scheduling on edge computing platforms with significantly reduced runtime complexity compared to the exact or brute-force scheduling algorithms. Moreover, the proposed RL model is generalizable to arbitrary DNN computational graphs within the same edge device domain without retraining. The contributions of this work can be summarized as follows: \textbf{(1)} to achieve the near-optimal scheduling performance with better scalability, we propose a novel RL-based scheduling framework that imitates the algorithmic behaviors of existing optimal scheduling algorithm, with data-independent training setups, i.e., training conducted fully on synthetic datasets; \textbf{(2)} the experimental results are conducted on a physical multi-stage pipelined Edge TPU system \cite{boroumand2021mitigating,yazdanbakhsh2021evaluation} via USB 3.0 communication interface which demonstrates significant Edge TPU inference runtime speedups over the commercial Edge TPU compiler with ten popular ImageNet DNNs models. Simultaneously, RESPECT offers up to $683 \times$ speedups in scheduling solving time, and up to {$2.5 \times$} performance speedup; and \textbf{(3)} finally, we compare the proposed RL method to the exact optimal scheduling method, where the proposed RL approach offers up to {{930$\times$} speedups for scheduling, with a near-optimal Edge TPU inference runtime performance}. Our results and Edge TPU deployment infrastructure are released publicly \footnote{\url{https://github.com/Yu-Utah/RESPECT}}. 
 
\section{Background}\label{sec:background}

In the modern deep learning frameworks, machine learning algorithms are represented as computational graphs, where each graph is a directed graph $G(V,E)$ with nodes $V$ describing operations, and edges $E$ representing the dataflows that connect the operators and input/output tensors. Practically, when deploying DNNs on hardware accelerators, the computational graphs are mostly represented as directed acyclic graphs (DAG), where the acyclic paths are unrolled to maximize the hardware performance. The computational graphs are mostly generated with static compilation.  

Specifically, the optimization objectives of scheduling computational graphs can be defined as follows: \textbf{Given:} \textit{(1)} A DAG $G(V,E)$ where $V$ represents the set of operations in the DNN computational graphs, and $E$ represents the set of edges; \textit{(2)} A set of scheduling constraints, which may include dependency constraints, resource constraints, execution time, memory allocations, etc. \textbf{Objective:} Construct an exact optimal schedule $S$ = {$s_0$,$s_1$,...$s_n$}, $n \leq |V|$, where $n$ represents the number of scheduling stages. The operation set in the computational graph $V$ will be allocated to $S$ that satisfies all scheduling constraints. For example, in a multi-stage pipelined Edge TPU system in Figure \ref{fig:sys_flow}, the resulted schedule assigns computation node $N_i$ to $s_0$ (Edge TPU:0), $N_k$ to $s_1$ (Edge TPU:1), $N_l$ and $N_o$ to $s_2$ (Edge TPU:2), $N_j$, $N_p$ and $N_q$ to $s_3$ (Edge TPU:3), etc.

Inference performance on edge devices is highly sensitive to scheduling solutions of computational graphs, due to the limited compute elements, memory, and communication bandwidth. The optimization problem of computational graph scheduling on edge devices is similar to the classic resource-constrained scheduling (RCS) problem, which has been widely studied. These studies have resulted in a set of heuristic scheduling methods (Hu's Algorithm), List Scheduling, and Force-Directed Scheduling \cite{paulin1989force,yang1993list}. It is believed that Google Edge TPU compiler adopts the heuristic methods to schedule computation graphs with an acceptable schedule solving time. Although heuristic methods can easily obtain the solution, the scheduling solutions could be far from optimum. Another viable scheduling method is iterative metaheuristics, such as simulated annealing, ant colony, and dynamic programming optimizations~\cite{micheli1994synthesis}. For example, \cite{ahn2020ordering} proposed dynamic programming based adaptive budgeting scheduling technique, which can optimize scheduling efficiently without either performance guarantees or significant long optimization runtime requirements.

On the other hand, the RCS problem can be solved with exact optimal methods using modern formal method solvers such as SAT and SMT solvers, or constraint solving such as Integer Linear Programming (ILP), after it maps to a constraint satisfaction problem which consists of logical connectives of linear constraints \cite{leiserson1991retiming,zhang2013sdc,yin2022exact}. It integrates specialized solvers with propositional satisfiability search techniques to achieve conflict-driven learning~\cite{de2011satisfiability}. Even though, such exact methods can produce the optimal solution, it has a drawback of high schedule solving time, making it difficult to apply to super large-scale computation graph scheduling. Thus, existing RCS scheduling methods have a clear trade-off between scheduling solving time complexity and the scheduling solution quality. This work aims to push the Pareto-frontier of traditional scheduling methods specifically for ML inference on edge devices, to offer near-optimum scheduling solutions at the optimization cost of fast heuristics. 

\section{Approach}\label{sec:approach}

Figure \ref{fig:sys_flow} shows the overview of RESPECT framework -- Step~\circled{1}: RESPECT takes the real-world DNN model as input and converts it to a hypergraph depending on the computation nodes and dependency between nodes. Step~\circled{2}: The embedding of the computation graph is generated in RESPECT, which includes node attributes (topological level, node's ID), node dependency (parents topo level, parents' IDs), and memory consumption (node's memory). 
Step~\circled{3}: A LSTM-PtrNet trained with reinforcement learning (RL) is employed to generate the schedule sequence with the embedded input queue $q$ from step~\circled{2}. Note that the LSTM-PtrNet model is trained with synthetic datasets to save the efforts for real-world data collection. 
Step~\circled{4}: The model is ready to be deployed to devices with the scheduling sequence from step~\circled{3}. Additionally, post-inference processing with domain-specific constraints is applied to make the deployment compatible with physical devices. For example, RESPECT will quantize the Tensorflow Model and deploy sub-models to real-world hardware devices using the Tensorflow Toco converter.

\subsection{DNN computational Graph Embedding}

\begin{figure*}
    \centering
\begin{subfigure}[b]{0.49\textwidth}
    \centering
   \includegraphics[width=1\linewidth]{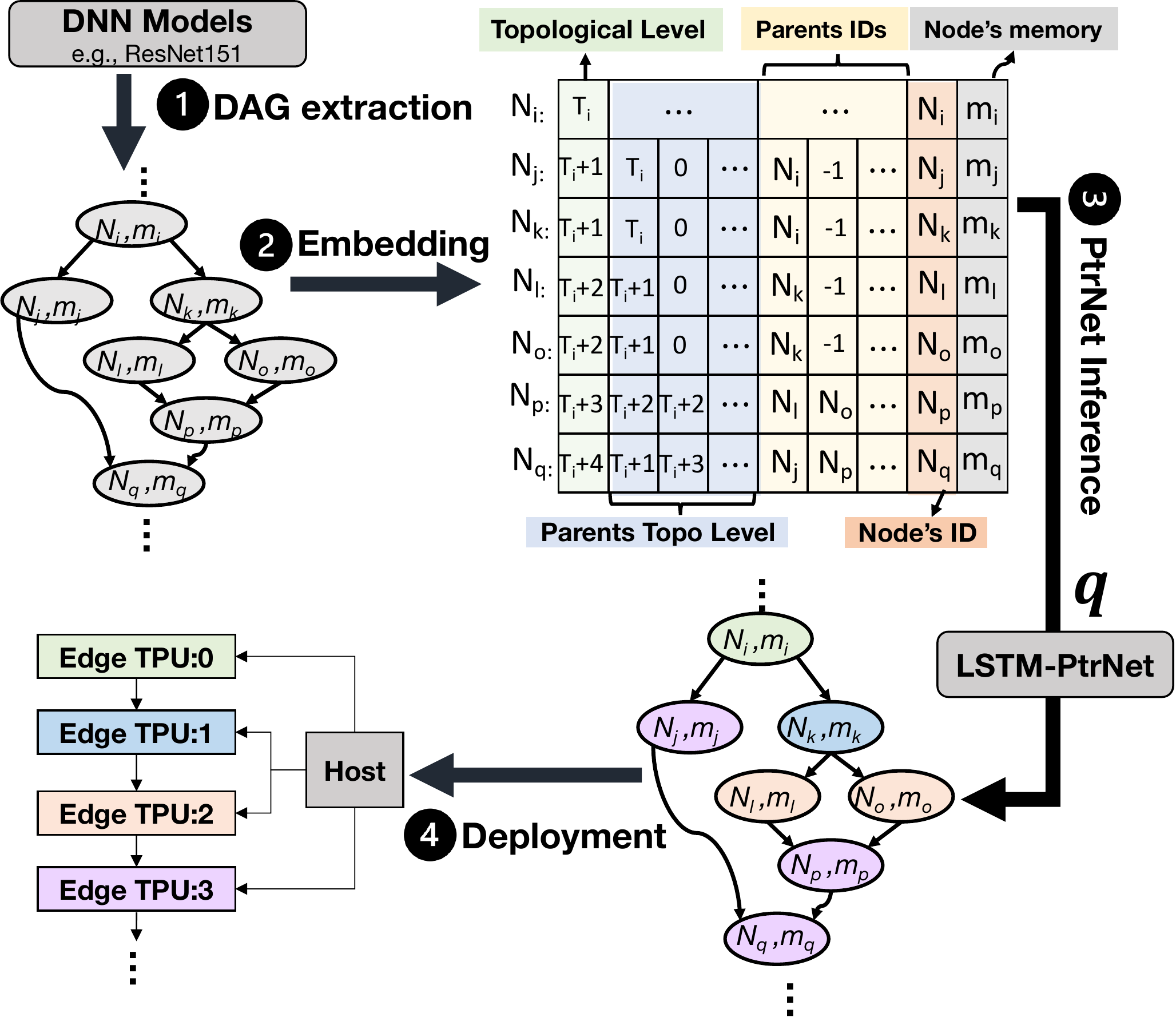}
    \caption{RESPECT framework overview -- Step \circled{1}: real-world DNN model DAG extraction; Step~\circled{2}: graph embedding based on topological level, node's ID and memory consumption; Step~\circled{3}: inference through the encoder-decoder based LSTM-PtrNet graph scheduler; Step~\circled{4}: hardware deployment.}
    \label{fig:sys_flow}
\end{subfigure}
\hfill
 \begin{subfigure}[b]{0.48\textwidth}
    \centering
    \includegraphics[width=0.8\linewidth]{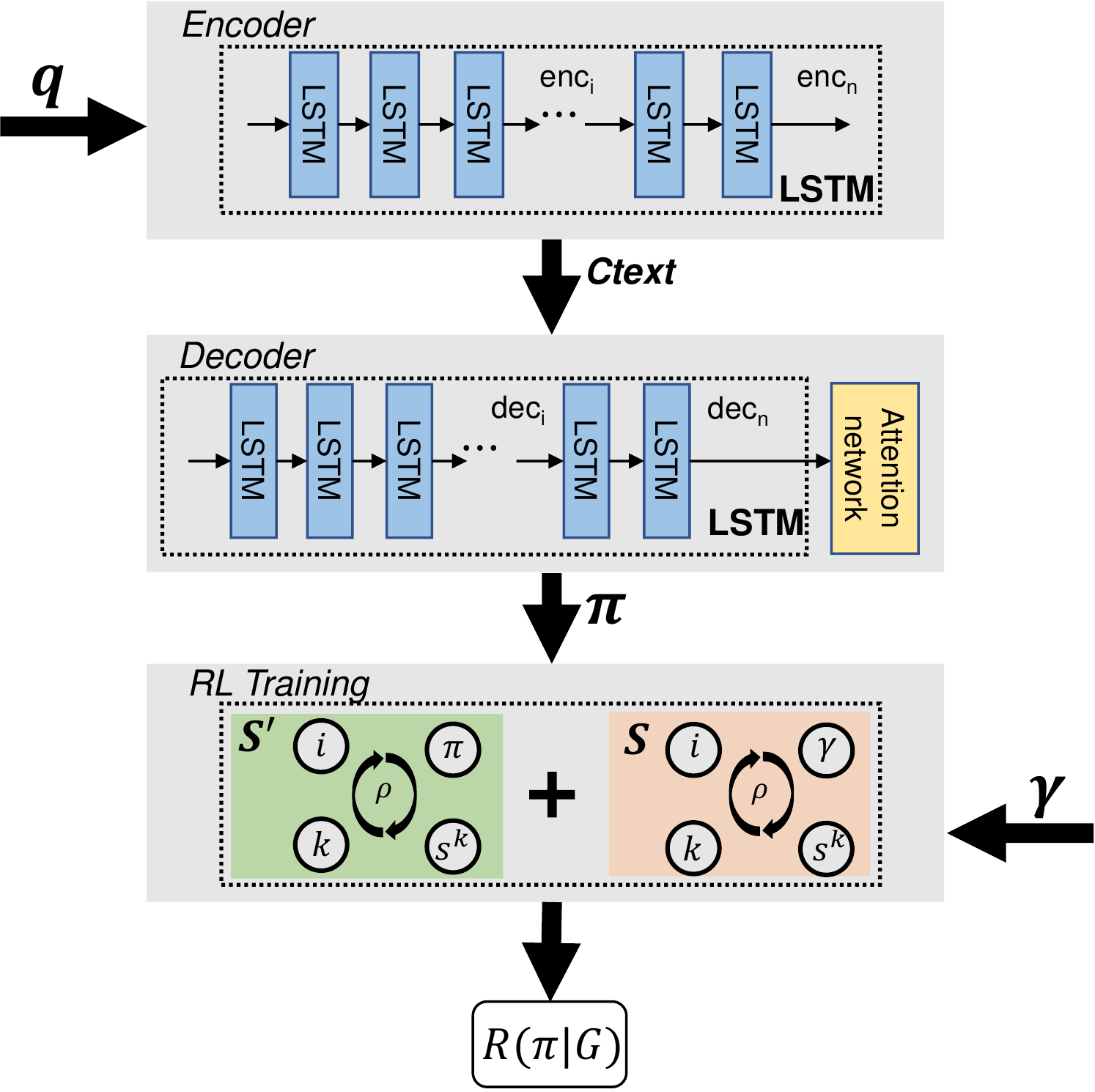}
    \caption{Detailed architecture of the LSTM-PtrNet model. The encoder network will digest the embedded input queue $q$ and transform it into a sequence of context $Ctext$, and produce a sequence of latent memory states $enc_i$. The decoder network will produce the predicted sequence $\pi$, and generate the latent memory state $dec_i$. RL model training is based on sequence $\pi$ and ground truth label $\gamma$.}
    \label{fig:sys_encoder}
\end{subfigure}
\caption{Overview of our RL-based scheduling framework.}

\label{fig:sys_overview}
\end{figure*}



For learning purposes, the computational graph is usually embedded into vector space. Considering the performance of scheduling on edge devices, the embedding of computational graph consists of four components as shown in Figure \ref{fig:sys_flow}: \textbf{(1)} \textit{absolute coordinates} generated based on topological ordering. For example, for node $N_i$, its absolute coordinates will be $T_i$ in the first column in its embedding. Note that there are many variants of topological orders of a given DAG. In this work, we use As-Soon-As-Possible (ASAP) ordering, where each node is ordered as close to the source node as possible; \textbf{(2)} \textit{relative coordinates} consisting of parent nodes absolute coordinates and parents IDs. For example, for node $N_j$, the relative coordinates will be its parents topological level $T_i$ and the parents IDs $N_i$. For source nodes, absolute coordinates are set to $0$, and their IDs are set to $-1$; \textbf{(3)} node IDs are unique integers generated by hashing all the operator names. For node $N_i$, its node's ID will be $N_i$; \textbf{(4)} cache and memory locality are one of the most critical metrics in DNNs acceleration. Thus, we add the memory consumption of each operator in the last embedding column. Specifically, absolute coordinates embedding represents the topological order of every node which mimics the feasible scheduling space for a given node. Relative positions are encoded using both the absolute coordinates and node IDs. Intuitively, relative coordinates maintain the dependency constraints and absolute coordinates. For example, in Figure \ref{fig:sys_flow}, the data dependency constraints for $N_p$ are the complete scheduling of its parent nodes, i.e., $N_l$ and $N_o$. Meanwhile, $N_j$ and $N_p$ can be scheduled together on available computation resources as there is no data dependency between them. Besides, the absolute coordinates of node $N_j$ and its parent node $N_i$ are $2$ and $1$, respectively. Thus, the valid schedule stages for $N_j$ can be calculated, i.e., any available stage between the absolute coordinates of $N_i$ and $N_j$. When $N_i$ is deployed to $s_0$, available stages for deploying $N_j$ include $\{s_0, s_1, s_2, s_3\}$. 


\subsection{Formulations and Neural Architecture}\label{sec:Arch and Conf}


As discussed earlier, brute-force and exact methods scheduling algorithms can generate the best scheduling solution search \cite{leiserson1991retiming,zhang2013sdc,CoZh06:sdc} at the cost of high runtime complexity, which fails in scaling up to large problems. While heuristic algorithms optimize the schedules more efficiently at scale, they suffer from the quality of results. Thus, we aim to develop an RL framework that imitates the algorithmic behaviors of any optimal scheduling algorithm (e.g., exact or brute-force), such that it performs polynomial time scheduling with near-optimum results at inference runtime.

\noindent
\textbf{RL Formulation for DNN computational Graph Scheduling} -- Given a computational graph DAG $G(V,E)$, the RL-agent is trained to develop a policy $\pi$, which picks computation nodes in the same order as a target exact or brute-force algorithm does. In this work, we define a sequence order generated by our method $\pi(i), i\le |V|$, and a sequence order figured by a given deterministic exact scheduling method as $\gamma(i), i\le |V|$. This exact scheduling method can be formulated using ILP, which offers the exact optimal schedule w.r.t a given optimization objective. The reward function is designed as the similarity comparison between the two sequences:
\begin{equation}
\begin{split}
R &= ~\frac{\sum (\pi(i)\cdot \gamma(i))}{\max (\sqrt{\sum \pi(i)^2}\cdot \sqrt{\sum \gamma(i)^2}, \epsilon)}
\end{split}
\label{equ:reward_simi}
\end{equation}
where $\epsilon$ is a small constant number to guarantee the validity of the equation.
Let $N_i\xrightarrow[]{} s_k, i\in [1, |V|]$, $k\in[0, n]$ be the solution of scheduling node $N_i$ at $s_k$. Let $S'=\{s_0,s_1,...s_n\}$, where $n$ is the number of Edge TPU stages, be the sequence produced by our policy $\pi(i)$, i.e., the output of the RL agent for a given computational graph. The targeted scheduling solution generated by the exact scheduling method is the ground truth label sequence, denoted as $S$. $S'$ and $S$ can be computed as follows:
\begin{equation}
\begin{split}
S' = \rho (\pi(i), s_k); ~~~~S = \rho(\gamma(i), s_k), i\in [1, |V|], k\in [0, n]
\end{split}
\end{equation}
where $\rho$ denotes the scheduling algorithm w.r.t the specific Edge TPU. Hence, {to imitate the behaviors of the targeted exact scheduling method}, the reward is designed using \texttt{cosine similarity} between the generated schedule sequence $S'$ and the ground truth sequence $S$ (Equation \ref{eq:reward}). 
\begin{equation}
\label{eq:reward}
\begin{split}
\small
R &= ~\frac{\sum (S_i^k\cdot {S'}_i^k)}{\max (\sqrt{\sum {S_i^k}^2}\cdot \sqrt{\sum {{S'}_i^k}^2}, \epsilon)}, i\in [1, |V|], k\in [0, n]
\end{split}
\end{equation}
 
While maximizing the reward function $R$, parameters of the stochastic policy $p(\pi | G)$ will be optimized to assign high probabilities to sequence order closer to the target sequence. The chain rule utilized to factorize the sequence probability distribution can be expressed as:
\begin{equation}
\small
\begin{split}
p(\pi | G) &= \prod_{i=1}^{|V|} p(\pi(i) | \pi(< i), G)
\end{split}
\end{equation}
where $\pi(< i)$ represents the output sequence generated by previous epochs. At each epoch, $p(\pi | G)$ is updated based on input computation graph and the previous sequence $\pi(< i)$.

\noindent
\textbf{RL Agent Architecture} -- We extend pointer network (PtrNet) architecture \cite{bello2016neural} as our RL agent. {PtrNet excels in finding the path with target objectives to be optimized. It achieves huge success in solving some combinatorial problems over graphs, such as the Traveling Salesman Problem \cite{bello2016neural}.} With the advantage of attention mechanism \cite{vaswani2017attention}, PtrNet reinforces the dependency constraints among nodes and overcomes the limitations of learning combinatorial graph algorithms with fixed size of graph inputs. 
Note that in order to fully evaluate the RL scheduling performance on physical edge devices, the results need to satisfy domain-specific hardware execution requirements, such as data dependency.

%
\begin{algorithm}[t]
\small
\SetAlgoLined
\textbf{Initialization}: $d \leftarrow$ decoder input; $C=\{Ctext_i\}_{i=1}^{|V|}$; $dec_0$; $\{emb_i\}_{i=1}^{|V|}$; $\theta, \omega, \beta$ (trainable parameters); \\
 \For{$i=1$ to $n$}{
      h, $dec_i$ $\xleftarrow[]{}$ Dec(d, $dec_{i-1}$);\\
      h $\xleftarrow[]{}$ glimpse(C$*\boldsymbol{\theta_g}$, $\boldsymbol{\omega_g}\cdot$h$+\boldsymbol{\beta_g}$);\\
      $P^i$ $\xleftarrow[]{}$ pointer($\tanh$(C$*\boldsymbol{\theta_p}$, $\boldsymbol{\omega_p}\cdot$ h + $\boldsymbol{\beta_p}$));\\
      $idx_i$ $\xleftarrow[]{}$ argmax$\frac{\exp{P_i^i}}{\sum_{j=1}^n \exp{P_j^i}}$;\\
      d $\xleftarrow[]{} emb_{idx_i}$}
 $S'$ $\xleftarrow[]{}$ $\rho$ ($\pi(i)$, $s_k$);~~S $\xleftarrow[]{}$ $\rho$($\gamma(i)$, $s_k$), $i\in [1, |V|]$, $k\in [0, n]$;\\
 
 \caption{Pointing Mechanism Decoding Flow}
 \label{Alg: Decodeing-flow}
\end{algorithm}

The RL agent architecture consists of encoding and decoding components, as shown in Figure \ref{fig:sys_encoder}. Specifically, each component is a Long Short-Term Memory (LSTM) cell.
\textbf{Encoder network} -- The encoder network digests the encoding input queue $q$ of nodes and transforms it into a sequence of context $\{Ctext\}_{i=1}^{|V|}$ representing the information of each node, where $Ctext_i\in R^d$, $d$ is the hidden dimension of LSTMs. The concatenation of all contexts will be used as reference matrix $C$ in picked nodes during decoding step, $C(i)=Ctext_i, i\in [1, |V|]$. When using LSTM as an encoder, it produces a sequence of latent memory states $\{enc_i\}_{i=1}^{|V|}$ recording the propagation encoding message from the first node to current one along with the contexts, where the final state $enc_{|V|}$ is the initial latent memory state for decoding. \textbf{Decoder network} -- When using LSTM as decoder network, it generates its own latent memory state $dec_i \in R^d$ at each decoding step $i$ to update the previous one. We illustrate the detailed decoding procedure in Algorithm \ref{Alg: Decodeing-flow}. With context reference matrix $C$ from encoder, decoder will produce a selection probability distribution over candidate computation nodes using pointing mechanism. 
The LSTM is employed as the decoder (Dec), which takes the node embedding $d$ and the last latent memory state $dec_{i-1}$ as inputs to produce the annotation $h$, and updates the latent memory state $dec_{i}$. Then attention networks including \textit{glimpse} and \textit{Ptr-Net} \cite{vinyals2015pointer} are employed to produce the sequence order $\pi$. Specifically, \textit{glimpse} updates $h$ with the context matrix $C$, and \textit{Ptr-Net} will produce the probability over candidates with the hidden state information from both encoder ($enc_{i}$) and decoder ($dec_{i}$). The next node will be selected with \texttt{softmax}.
Once a computation node $N_{idx_i}$ with the highest probability is picked, its embedding $emb_{idx_i}$ will be passed as the input $d$ to LSTMs together with the information of the last latent memory state $dec_{i-1}$ during the next decoding step. The logits of the nodes that appeared in the solution $\pi$ are set to $-\infty$ in the network to guarantee the validity of the solution. The input to the first decoding step $dec_0$ is a trainable parameter sharing the same dimension with node embedding.

\noindent
\textbf{RL Training} -- We use the model-free policy-based RL training method to optimize the parameters of a pointer network denoted among $\theta$. The learning objective is the expected similarity of node distribution. Given an input graph $G$, the optimization objective is to maximize the following cosine similarity reward:
\begin{equation}
\label{eq:opt}
\begin{split}
J(\theta | G) &= \mathbb{E}_{\pi\sim p_{\theta}(\cdot | G)}(1-R(\pi | G))
\end{split}
\end{equation}
{In this work, we deploy policy gradient algorithms and stochastic gradient descent algorithms to optimize the parameters \cite{williams1992simple}. Reward as a score to evaluate the resulted distribution probability is applied as a coefficient in gradient calculation, such that the gradient of Equation \ref{eq:opt} is constructed as follows}: 
\begin{equation}
\small
\begin{split}
\nabla_{\theta} J(\theta | G) &= \mathbb{E}_{\pi\sim p_{\theta}(\cdot | G)}[(1-R(\pi | G)-b(G))\nabla_{\theta}\log p_{\theta}(\pi | G)]
\end{split}
\end{equation}
where $b(G)$ represents a baseline in LSTM-PtrNet model that performs the best among all the past training iterations, i.e., rollout baseline \cite{kool2018attention}, and it is applied in loss function to reduce the variance of the gradients. 
Note that, unlike existing RL-based scheduling works, our framework imitates the behaviors of any scheduling algorithm, such that the resulted RL agent can perform as generally as the imitated deterministic algorithm.

\noindent
\textbf{Synthetic training dataset} -- 
The limited number of computation graphs for training the RL models restricts the generalizability of RL in solving graph-based combinatorial optimization problems. Besides, to simultaneously minimize the training efforts, the training dataset needs to balance data coverage and graph size (i.e., $|V|$ and $|E|$ in $G(V,E)$). There is a critical advantage to use synthetic data sampling during RL training: synthetic graph sampler has full control over graph complexity, and memory attributes of operators, which offers much better data coverage over real-world DNN computation graphs. Therefore, we use the synthetic graphs only in training, where we integrate a DAG sampler into our RL training framework which randomly generates network graphs with $|V| = 30$ but with different graph complexities. The synthetic datasets are designed to mimic the structures and properties of DNN computational graphs. The complexities of the trained graphs can vary with different graph degrees ($deg(V) \in \{2, 3, 4, 5, 6\}$), which is the maximum number of incoming edges connecting to the vertices $V$ in graph $G(V, E)$. RESPECT is trained with 1 million random graphs, with 200,000 graphs for each $deg(V)$ in \{2, 3, 4, 5, 6\}. 

\noindent
\textbf{Post-Inference Processing} -- Unlike TSP algorithms, DNN computational graph scheduling needs to satisfy domain-specific constraints to successfully deploy the scheduled graphs on hardware platforms. {Thus, a post-inference processing procedure is added to our RL framework, which is executed at the deployment stage, and takes the inference output $S'$ as input.} To be specific for our evaluation on Edge TPU platforms, this procedure corrects the dependency violation by simply pushing the involved node forward, which is a deterministic step with minimum changes to the RL solution. Besides, Edge TPU hardware requires children nodes of any node to be in the same pipeline, where the post-inference procedure assigns these nodes to the earliest predicted stage.

\section{EXPERIMENTAL RESULTS}\label{sec:results}

 \begin{figure}[!htb]
     \centering
     \includegraphics[width=0.35\textwidth]{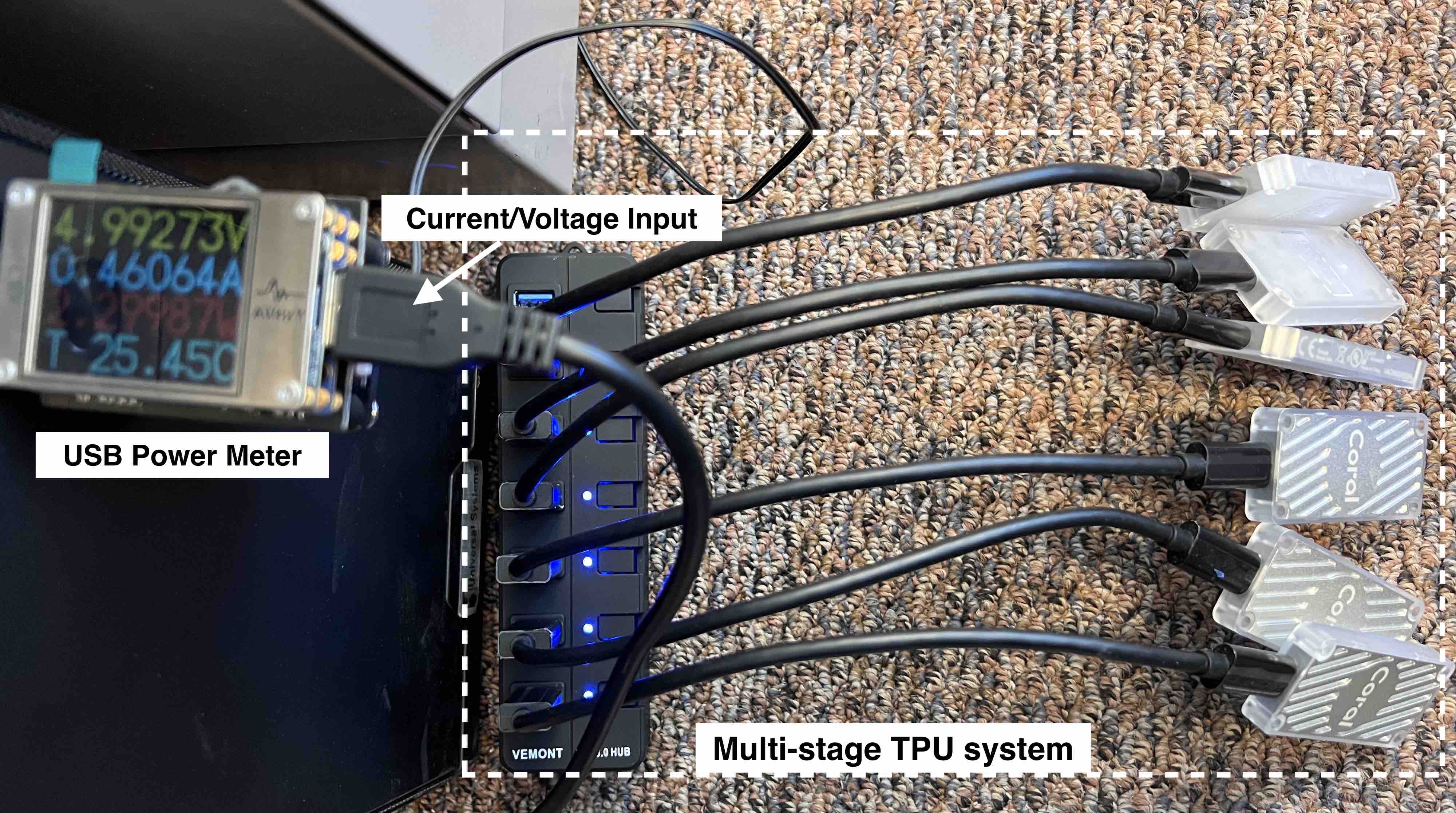}
     \caption{{Illustration of our multi-stage pipeline Edge TPUs and energy efficiency evaluation system via USB 3.0 communication interface.}}
     \label{fig:power_measure_sys}
 \end{figure}
 

\begin{figure*}[!htb]
    \centering
    \includegraphics[width=1\linewidth]{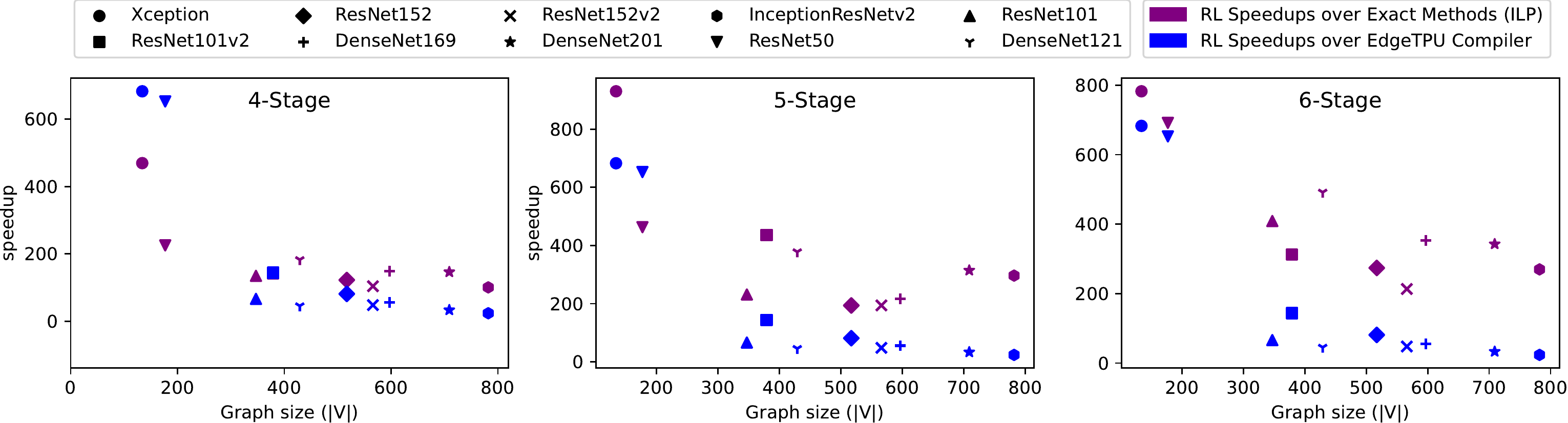}
    \caption{Schedule solving time across ten real-world ImageNet models with different graph sizes. The horizontal-axis and vertical-axis represent graph sizes $\lvert V \rvert$, and speedups over the baselines, respectively. RL method offers $24-683\times$ speedups over Edge TPU compiler and $100-930\times$ over exact methods.}
    \label{fig:scheduling_overhead}
\end{figure*}

In this section, we aim to experimentally demonstrate the effectiveness of the proposed RL-based scheduling approach which offers near-optimal scheduling solutions with runtime cost of inference, using real-world physical Edge TPU devices. The two comparison baselines include commercial Edge TPU compilers (e.g., Google's Edge TPU compiler\footnote{\url{https://coral.ai/docs/edgetpu/compiler/}}), which adopts the heuristic method to schedule computation graphs, and an exact-optimal scheduling method conducted on constraint solving scheduling using ILP solver \cite{leiserson1991retiming,zhang2013sdc}. We evaluate RESPECT in three aspects: 1) on-chip inference runtime, 2) solving runtime, and 3) the optimization gaps to exact optimal solutions.

Our framework integrates RESPECT scheduling with Tensorflow-Lite (TFLite) and Edge TPU Compiler to complete the deployment flow. It takes single or multiple DNN models and the number of pipeline stages as inputs, and outputs $n$ partitioned subgraphs for deployment on Edge TPU devices. The framework involves graph construction, solving via RESPECT, solution extraction, and deployment. 

\noindent
\textbf{Experiment and evaluation setups} -- 
We use \textbf{ten} popular ImageNet classification models {(Table \ref{tbl:dnn_graphs})} as the benchmarks. 
Training, inference, and explorations of the proposed RL methods are conducted on one Nvidia 2080 Ti with Intel Xeon Gold 6230 x20 CPUs. 
For the exact method, the optimization ILP formulations are solved using IBM ILOG CPLEX. The scheduling solution with heuristic method is acquired from Google's Edge TPU compiler$^1$. RESPECT is implemented with PyTorch. 
The neural architecture encoder and decoder are built with LSTMs with $256$ cells. Training is conducted on 300 epochs with the learning rate of $10^{-4}$ and batch size of $128$ using Adam optimizer. Besides, we build a central-hosted EdgeTPU system for experiment results evaluation which is shown in Figure \ref{fig:power_measure_sys}.

\begin{figure}[!htb]
  \centering
  \begin{minipage}{0.5\textwidth}
    \centering
    \subcaptionbox{4-stage}
      {\includegraphics[width=1\linewidth]{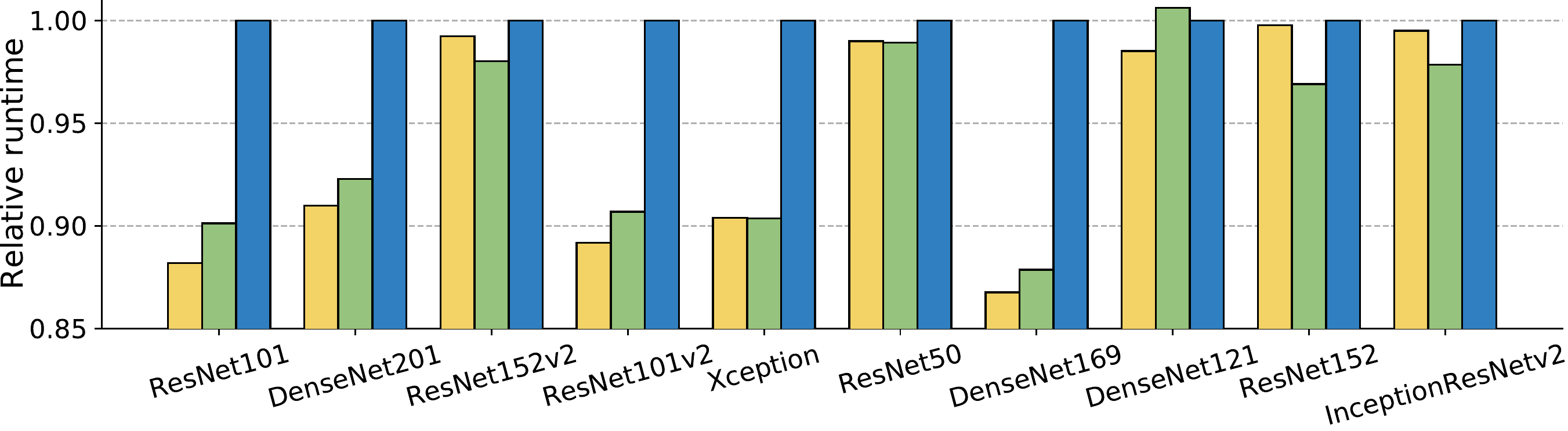}}
    \subcaptionbox{5-stage}
      {\includegraphics[width=1\linewidth]{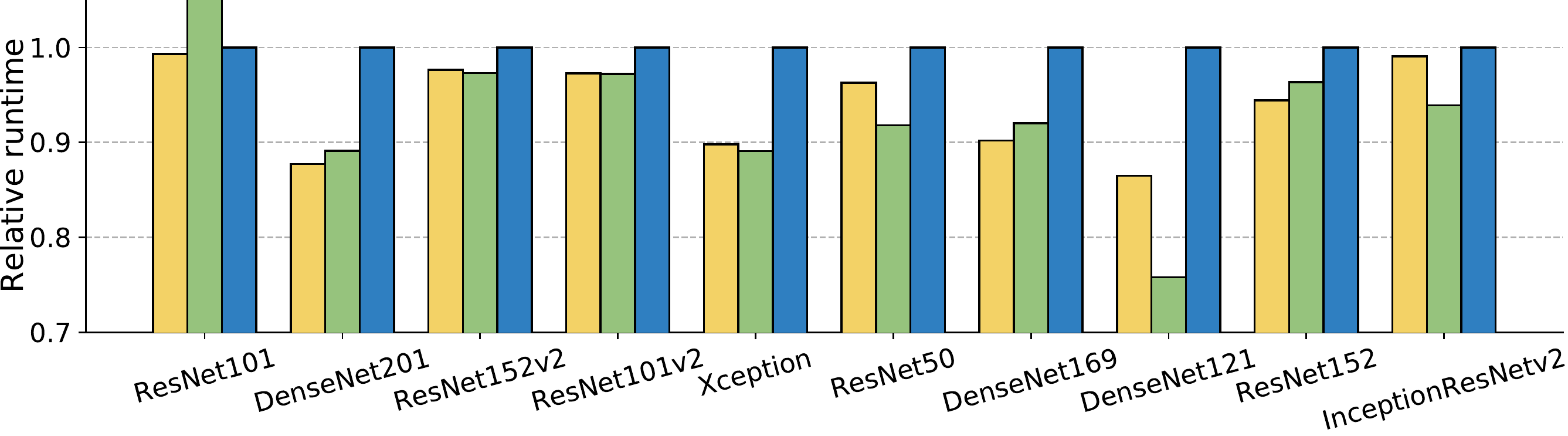}}
    \subcaptionbox{6-stage}
      {\includegraphics[width=1\linewidth]{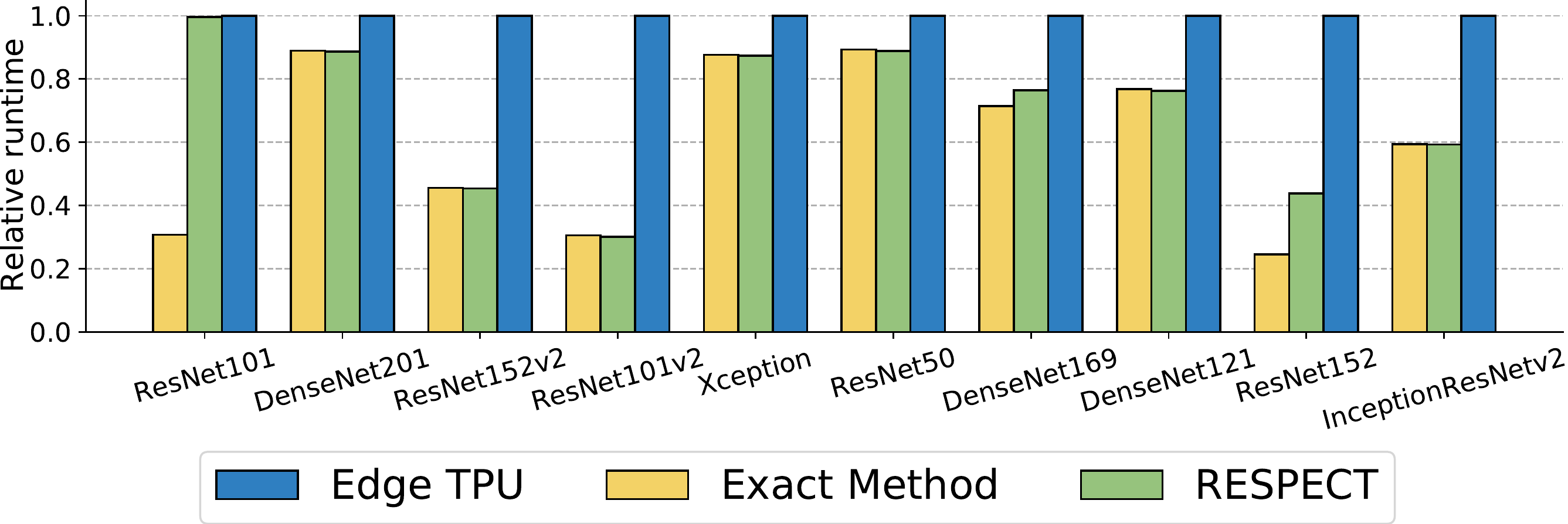}}
  \end{minipage}
  \caption{{Multi-stage pipelined Edge TPUs inference runtime comparisons between the proposed RL methods, exact methods, and commercial Edge TPU compiler (baseline scale=1).} RL methods provide a near-optimum scheduling assignment. The runtime performance has been consistently improved over commercial compilers with 4-, 5-, and 6-stage pipelined Edge TPU systems (e.g., ResNet101v2 and ResNet152 execute $2.5\times$ faster than the Edge TPU compiler).}
  \label{fig:edgetpu_runtime}
  
\end{figure}

\begin{table}[!h]
\scriptsize
    \captionof{table}{Statistics of DNNs models and their computational graphs used for evaluating inference runtime on Pipelined Edge TPU Systems.}
    \label{tbl:dnn_graphs}
    \resizebox{0.45\textwidth}{!}{%
    \begin{tabular}{|c|c|c|c|c|}\hline
          & \begin{tabular}{@{}c@{}} Xception \\ 
          \end{tabular} & \begin{tabular}{@{}c@{}} ResNet50 \\ 
          \end{tabular} & \begin{tabular}{@{}c@{}} ResNet101 \\ 
          \end{tabular} & \begin{tabular}{@{}c@{}} ResNet152 \\ 
          \end{tabular} \\\hline
        $\mathbf{|V|}$ & 134 & 177 & 347 & 517\\\hline
        $\mathbf{deg(V)}$ & 2 & 2 & 2 & 2\\\hline
        \textbf{Depth} & 125 & 168 & 338 & 508\\\hline
         & \begin{tabular}{@{}c@{}} DenseNet121 \\ 
         \end{tabular} & \begin{tabular}{@{}c@{}} ResNet101v2 \\ 
         \end{tabular} & \begin{tabular}{@{}c@{}} ResNet152v2 \\ 
         \end{tabular} & \begin{tabular}{@{}c@{}} DenseNet169 \\ 
         \end{tabular} \\\hline
        $\mathbf{|V|}$ & 429 & 379 & 566 & 597\\\hline
        $\mathbf{deg(V)}$ & 2 & 2 & 2 & 2\\\hline
        \textbf{Depth} & 428 & 371 & 558 & 596 \\\hline
          & \begin{tabular}{@{}c@{}} DenseNet201 \\ 
          \end{tabular} & \begin{tabular}{@{}c@{}} InceptionResNetv2 \\ 
          \end{tabular} & & \\\hline
        $\mathbf{|V|}$ & 709 & 782 & & \\\hline
        $\mathbf{deg(V)}$ & 2 & 4 & & \\\hline
        \textbf{Depth} & 708 & 571 &  & \\\hline
    \end{tabular}}
    \label{DNNs}
    \vspace{-3mm}
\end{table}

\subsection{Edge TPU Inference Runtime Evaluation}\label{sec:edgetpu_runtime}



We compare the scheduling results of RESPECT with the heuristic method based Edge TPU compiler and exact methods based ILP solver. The results included in Figure \ref{fig:edgetpu_runtime} are the average runtime of 10 rounds of 1,000 ImageNet inferences. The vertical axis represents the inference runtime results normalized w.r.t the runtime obtained using Edge TPU compiler.

As shown in Figure \ref{fig:edgetpu_runtime}, the proposed RL scheduler consistently outperforms the commercial Edge TPU compiler. Specifically, the RL scheduler provides 1.06$\times$, 1.08$\times$, 1.65$\times$ speedups over Edge TPU compiler in 4-, 5-, and 6-stage pipelining options. For example, ResNet101v2 and ResNet152 execute 2.5$\times$ faster than Edge TPU compiler in 6-stage pipelining. Second, compared to the scheduling solutions generated with the exact optimal scheduling method solved by ILP, the proposed RL approach demonstrates near-optimum Edge TPU inference performance, but with a significantly reduced runtime in scheduling. Detailed discussions on solving time complexity are shown in Section \ref{sec:solv_scalability}. Third, the average Edge TPU inference runtime improvements of the proposed RL approach over Edge TPU compiler increase as the number of pipeline stages increases. For example, our proposed RL method provides a speedup of $1.03\times$ for ResNet152 in a 4-stage system, and the inference runtime speedup increases to $2.28\times$ in a 6-stage system. The main reason is that as the number of stages increases, the complexity of the scheduling optimization increases, where the performance of heuristics methods will further degrade.

{\textbf{Performance modeling miscorrelation} -- From the analysis of runtime performance shown in {Figures \ref{fig:edgetpu_runtime}}, we observe that exact methods sometimes perform worse than RESPECT. This is due to the miscorrelation between the high-level abstract modeling of computational graphs and the actual systolic hardware specifications, in which the hardware architecture and backend compilation are closed source. Such miscorrelations of the high-level performance modeling can be used to guide the scheduling optimizations but introduce optimization noise to the actual on-chip performance. Specifically, in this work, both the exact method and RESPECT optimize the DNN model scheduling from the aspects of the memory allocation and communication cost. Note that directly modeling or estimating the runtime from the computation graph and its schedule is impossible given the complexity of the Edge TPU compute components and caching/memory system \cite{boroumand2021mitigating}}.

\subsection{Solving Scalability Evaluation}
\label{sec:solv_scalability}

In this section, we further compare the schedule solving runtime of our proposed RL framework with the same set of baselines on the same set of DNN models.  
{The results are shown in Figure \ref{fig:scheduling_overhead}, where the {horizontal-axis} represents the graph sizes, i.e, $|V|$ in the computation graph $G(V, E)$. The {vertical-axis} shows the speedup in solving runtime of the RL framework over the two baselines.}



Three important conclusions can be summarized from Figure \ref{fig:scheduling_overhead}: \textbf{(1) Compared to the heuristic methods (Edge TPU compiler), the main advantage of the proposed RL approach is the improved quality of scheduling in Edge TPU inference runtime.} Specifically, our RL approach offers $24\times$ to $683\times$ speedups over Edge TPU compiler, in generating the scheduling solutions, with up to 2.5$\times$ inference runtime improvements discussed in Section \ref{sec:edgetpu_runtime}. For example, for ResNet152, the RL generated schedule offers 2.4$\times$ speedups in Edge TPU inference runtime and simultaneously improves the scheduling solving time by 85$\times$. \textbf{(2) Compared to the exact method conducted on constraint solving scheduling, our RL approach offers $100\times$ to $930\times$ speedups in generating scheduling solutions, where the Edge TPU inference runtime is almost identical}. For example, for the Xception model, the Edge TPU inference runtime is almost the same for RL approach and the exact method, while the RL scheduling solving time is 930$\times$ faster than the exact method. 




\subsection{Gap-to-Optimality Analysis}

\begin{figure}[!htb]
  \centering
  \begin{subfigure}[b]{0.5\textwidth}
  \centering
  \includegraphics[width=1\linewidth]{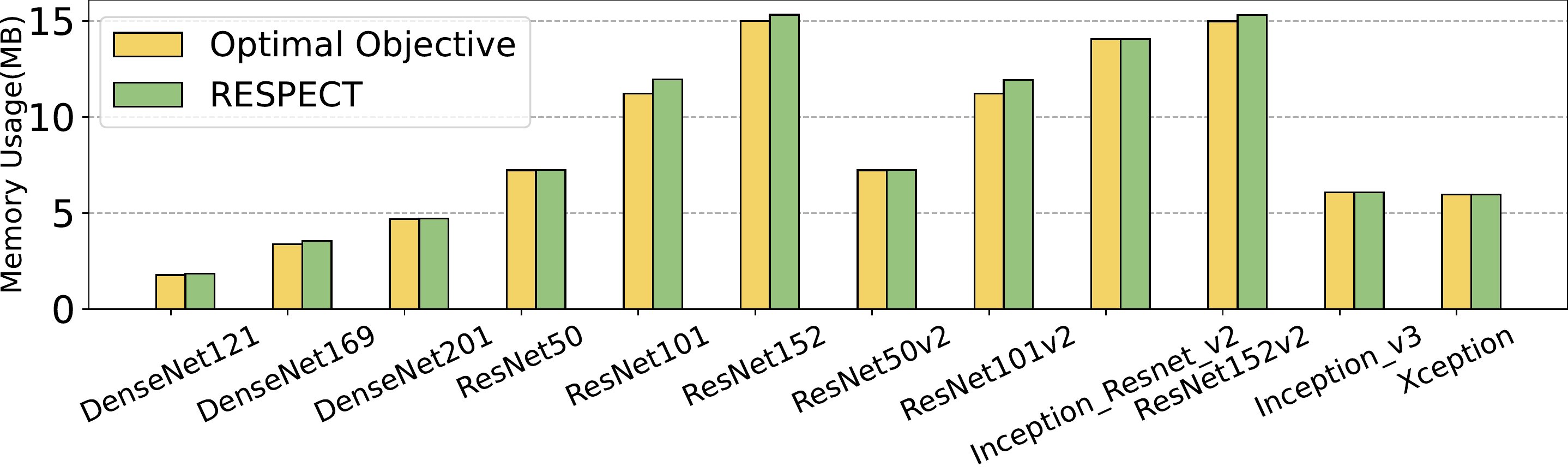}
  \caption{4-stage}
  \label{fig:gap_4_stage}
  \end{subfigure}
  \begin{subfigure}[b]{0.5\textwidth}
  \centering
  \includegraphics[width=1\linewidth]{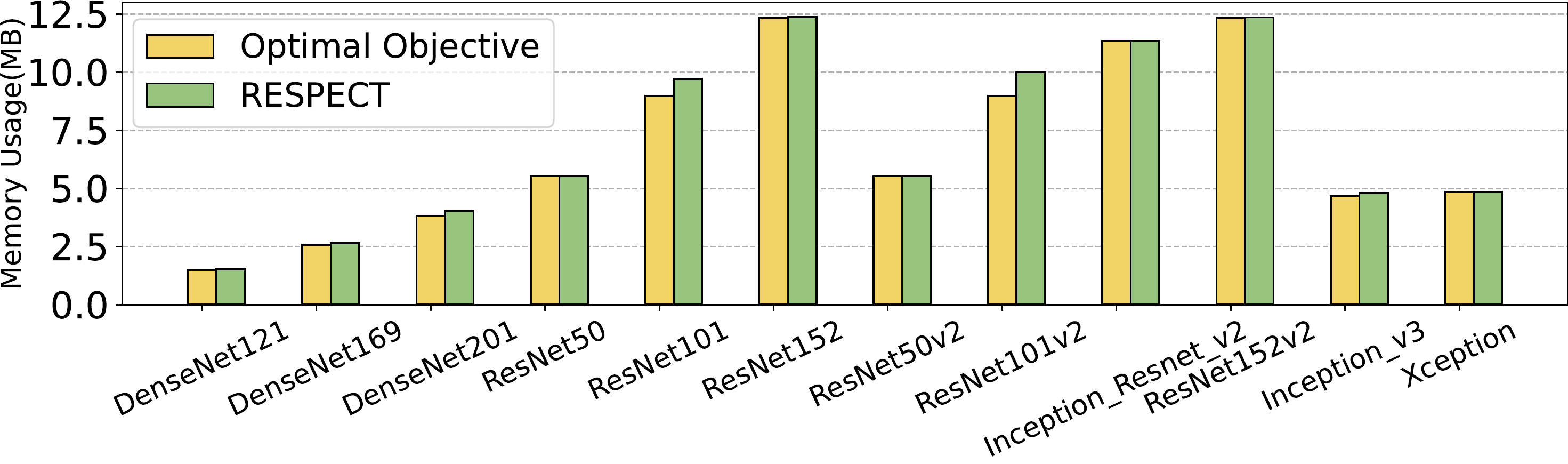}
  \caption{5-stage}
  \label{gap_5_stage}
  \label{fig:gap_5_stage}
  \end{subfigure}
  \begin{subfigure}[b]{0.5\textwidth}
  \centering
  \includegraphics[width=1\linewidth]{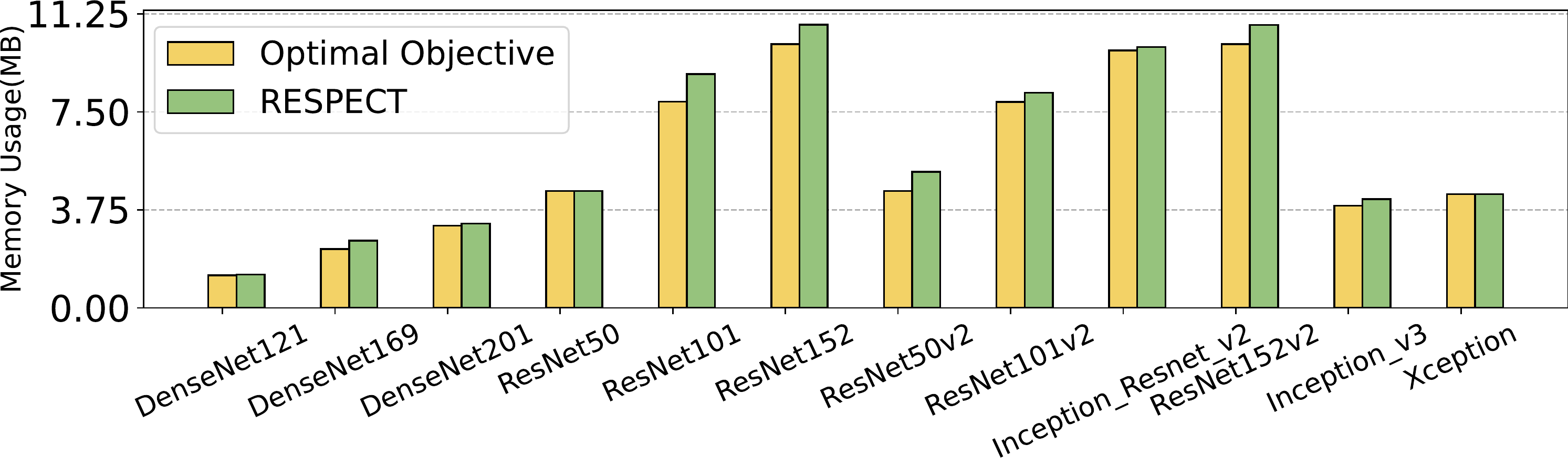}
  \caption{6-stage}
  \label{gap_6_stage}
  \label{fig:gap_6_stage}
  \end{subfigure}
  \caption{Gap-to-optimal analysis between RESPECT and exact scheduling. (a)--(c): Parameter caching results of ImageNet DNNs models cross 4-, 5-, 6-stage pipeline settings.}
  \label{fig:optimal_gap}
\end{figure}

Finally, we perform a generalizability analysis to demonstrate the capabilities of generating near-optimal scheduling solutions of the RESPECT framework. 
We confirm the generalizability of our RL framework by evaluating the following metrics: quality-of-results w.r.t the exact optimal solutions, namely \textit{gap-to-optimal} of RESPECT inference. Specifically, we collect the parameter caching values, including on-cache and off-cache memory caching values across all pipeline stages, and calculate the absolute differences in peak memory usage per stage between exact-optimal and RESPECT scheduling solutions. The results are conducted on the ImageNet DNNs models, shown in Figures \ref{fig:gap_4_stage} to \ref{fig:gap_6_stage}. We can see that RESPECT provides near-optimal results in parameter caching in comparison to exact optimal solutions, with only 2.26\%, 2.74\%, and 6.31\% average performance gap-to-optimal, across all 4-, 5-, and 6-stage pipeline settings.

\section{Conclusion}
This work presents a reinforcement learning based scheduling framework, which imitates the behaviors of optimal optimization algorithms in significantly reduced runtime complexity. Our framework has demonstrated up to $2.5\times$ runtime speedups over the commercial Edge TPU compiler, using ten popular ImageNet models on three different physical pipelined Google Edge TPU systems. More importantly, compared to the exact optimization methods solved by heuristics and brute-force, the proposed RL scheduling improves the scheduling solving time by several orders of magnitude.


\noindent
\textbf{Acknowledgment} -- This work is supported by National Science Foundation (NSF) under NSF-2047176, NSF-2019336, NSF-2008144, and NSF-2229562 awards.
 

\small
\bibliographystyle{IEEEtranS}
\bibliography{ref.bib,cds.bib}

\end{document}